# Surface Josephson plasma waves in a high-temperature superconductor

Qianbo Lu[1,2,3], Anthony. T. Bollinger[4], Xi He[4,5], Robert Sundling[4,6], Ivan Bozovic[2,4,5] and Adrian Gozar[1,2,7,*]

[1] Department of Applied Physics, Yale University, New Haven, CT 06520, USA
[2] Energy Sciences Institute, Yale University, West Haven, CT 06516, USA
[3] Shaanxi Institute of Flexible Electronics, Northwestern Polytechnical University, Xi'an, 710072, China
[4] Brookhaven National Laboratory, Upton, NY 11973, USA
[5] Department of Chemistry, Yale University, New Haven, CT 06520, USA
[6] Zensoft, Inc., Madison, Wisconsin 53705, USA
[7] Department of Physics, Yale University, New Haven, CT 06520, USA

## ABSTRACT

**Electron density oscillations with acoustic dispersions and sustained at boundaries between different media provide information about surface and interface properties of hetero-structures. In ultra-thin metallic films these plasmonic excitations are heavily damped. Superconductivity is predicted to reduce dissipation allowing detection of these resonances. Emerging low-loss interface Cooper-pair waves have been studied before, however, the observation of surface-confined Josephson plasmons has remained elusive. Here, we report on generation and coupling to these excitations in an ultrathin single-crystal film of high-temperature superconductor $La_{1.85}Sr_{0.15}CuO_4$. The film becomes brighter than Au below the critical temperature when probed with sub-gap THz photons. We show that the enhanced signal in the superconducting state, which can be visualized with a spatial resolution better than λ/3,000, originates from near-field coupling of light to surface Josephson plasmons. Our results open a path towards non-invasive investigation of enhanced superconductivity in artificial multilayers, buried interface states in topological hetero-structures, and non-linear phenomena in Josephson devices.**

## INTRODUCTION

Propagating sound-like collective modes in superfluids are converted to higher-energy plasma modes in superconductors[1]. Several decades ago, low-lying excitations with linear dispersions were detected in the microwave region, in aluminum films[2]. These are known as Carlson-Goldman modes. They consist of balanced oscillations of supercurrents and normal carriers[3] and are overdamped except in a narrow region close to $T_c$. In thin films and superlattices, a second type of superfluid acoustic mode should exist, where the electromagnetic field is confined to the interfaces between superconducting and dielectric regions[4]. Two main obstacles make the observation of these excitations difficult for traditional, far-field, optical techniques[5-7]. One is sensitivity, i.e. the capability to distinguish surface modes from bulk contributions. The other is generic to acoustic branches and entails overcoming the momentum mismatch between free-space photons and gapless plasmons. This can be realized in principle by coupling the evanescent waves through a prism, by periodic nano-fabricated arrays, or by bringing a nano-sized light source close to the sample[6,8,9]. Artificial periodic corrugation was indeed used to observe the emergence of microwave surface waves in superconductors[10-12]. These studies highlighted the role of reduced dissipation in the superfluid state and established these materials as candidates for temperature-tunable plasmonic structures[10,13]. In this work we use the second, scattering-based, approach, see Fig. 1, which has the benefit of increased spatial resolution due to the nanometer probe size.

There is also a practical interest in materials hosting stacks of conductive metallic sheets ordered with atomic scale precision as they can be used as building blocks for plasmonic devices. Along with doped semiconductor superlattices or modern two-dimensional (2D) materials, superconducting cuprates are examples of such layered systems[14-16]. Coupling mechanisms and the requirement of large momentum transfers for mapping dispersion branches gave electron energy loss spectroscopy the leading role in the study of plasmons, including pioneering work which enabled the extension of this concept to surfaces and interfaces[17]. Potential applications are met with substantial challenges because of electrical losses in conventional metallic structures[18]. The same problem is present and amplified in the normal state of superconductors with a high critical temperature ($T_c$). Topological protection was found to strongly suppress damping of acoustic plasmon excitations propagating on the surface of topological insulators[19,20]. Dissipation is also expected to be significantly reduced below $T_c$ in superconductors[4]. Optical probing of these materials in the sub-gap regime are of interest because below $T_c$ they are predicted to support low-loss plasmon waves and confinement of the photon field at deep sub-wavelength scales[13]. Near-field optics brings in the additional advantage of higher spatial resolution along with sub-surface sensitivity[21]. For high-$T_c$ cuprates, the inherent non-linearity of the Josephson coupling between the superconducting $CuO_2$ planes may enable photonic applications exploiting light-matter interactions in the THz range[22], with superfluid surface plasmon excitations predicted to play a special role[7,23,24]. The main result of this work is the observation of these modes in strongly anisotropic single crystal superconductors, where their energy and dispersions depend crucially on the Josephson coupling between adjacent layers.

Systems of coupled metallic layers harbor a variety of plasmonic branches and here we describe their salient features (see also Supplementary Note 4 and Supplementary Figure 3). There is a very large energy cost associated with excitations of plasmons in bulk metals. If the same materials are confined to 2D, this energy becomes vanishingly small in the long-wavelength limit; the oscillation frequency scales as $\omega(q \to 0) \propto q^{1/2}$, where $q$ is the in-plane momentum[25]. Quasi-2D surface excitations confined to planar interfaces between media with dielectric permittivities of opposite sign are also gapless but they display a sound-like dispersion $\omega(q \to 0) \propto q$, see for example ref. [9]. Interestingly, a change from the expected square root to linear dispersion for 2D electronic states is also generated by bulk-crystal screening effects: examples are the acoustic surface plasmons in metals such as those in Be(0001), ref.[26]. A cross-over between these two regimes can be achieved as well in a configuration where the 2D electron gas is screened by the proximity of a conductive planar electrode/gate[12], or by stacking 2D conductive and insulating layers in alternation on top of one another, ref. [27]. The latter geometry exhibits the (quasi-)2D to 3D panoply of gapped or gapless plasmon dispersions $\omega(q, k_z)$ as dictated by $k_z$, the momentum perpendicular to the planes[28]. In addition to these propagating guided modes, superlattices also display their characteristic surface-confined acoustic modes, where the electric and magnetic fields decay exponentially with the distance both inside and outside the layered material[6,29].

In high-$T_c$ copper oxides the normal state transport is extremely anisotropic: it is metallic in the *ab*-plane (parallel to the CuO$_2$ layers) and insulating/incoherent in the perpendicular, *c*-axis, direction. Below $T_c$, copper oxides behave like a series array of intrinsic Josephson junctions[14]. Tunneling between CuO$_2$ planes allows propagation of bulk plasmons only above $\omega_J$, the screened Josephson plasma energy [6,8,23]. Josephson plasma edge can be probed with far-field techniques and its measured value ranges from $\omega_J$ ~50 GHz in Ba$_2$Sr$_2$CaCu$_2$O$_8$ (ref. [30]) to ~1.5 THz in La$_{2-x}$Sr$_x$CuO$_4$ (ref. [31]) and even higher in YBa$_2$Cu$_4$O$_8$ (ref. [32]). Surface Josephson plasma waves (SJPWs) remain the only excitations with acoustic dispersion in these materials. An anisotropic two-fluid model was shown to capture the dielectric properties of the copper oxides[33]:

$$\varepsilon_\alpha(\omega) = \varepsilon_{\infty,\alpha} - \frac{\omega_{pS,\alpha}^2}{\omega^2} - \frac{\omega_{pN,\alpha}^2}{\omega^2 + i\Gamma\omega}, \qquad \alpha = ab, c \qquad (1)$$

Here '*S*' and '*N*' stand for the superfluid and normal components, $\omega_p$ is the plasma frequency (for the *c*-axis direction $\omega_J = \omega_{pS,c}/\sqrt{\varepsilon_{\infty,c}}$) and $\Gamma$ is the scattering rate. The first term in Eq. (1) follows from the London model and the second is a Drude-model description of the uncondensed carriers. In optimally doped La$_{1.85}$Sr$_{0.15}$CuO$_4$ (LSCO) the ratio of the in-plane to the *c*-axis superfluid plasma frequencies is $\omega_{pS,c}/\omega_{pS,ab} \approx 25$ at low temperatures[31,34].

**RESULTS**

**Film synthesis and near-field data**

We achieved coupling to SJPWs in a 13 nm thick LSCO film by using a customized cryogenic system based on Atomic Force Microscopy combined with Scanning Near-Field Optical Microscopy[35,36] (AFM-SNOM), see Fig. 1, Methods and Supplementary Note 1. For film synthesis, we have used an advanced Atomic-Layer-by-Layer Molecular Beam Epitaxy (ALL-MBE) system, a technique proven to provide the highest-quality, single-crystal LSCO films with atomically smooth surfaces and interfaces[37]. The surface morphology and crystalline structure of the film were monitored by reflection high-energy electron diffraction, in real time. A source of pure ozone was used to ensure sufficient oxidation under high-vacuum conditions. Single-crystal LaSrAlO$_4$ substrates were polished perpendicular to the crystallographic [001] direction. The films are epitaxially locked to the substrate and pseudomorphic; the CuO$_2$ layers are parallel to the LSCO film surface. During the growth, we kept the ozone partial pressure at $p = 2 \times 10^{-5}$ Torr and substrate temperature at $T_s = 650$ ºC. To protect the film surface, we cover the films *in situ* with 10 nm thick layer of gold, deposited at room temperature. The device shown in the inset of Fig. 1b was patterned using photolithography and Argon-ion milling the LSCO film into a 10 mm long, 20 µm wide strip with 64 lateral leads to large contact pads. The protective gold layer was removed from the active LSCO strip by an appropriate gold etch.

The AFM-SNOM system was optimized for laser light access and large collection throughput, as well as for AFM tip positioning and device visualization[38], see Fig. 1. All data shown here were acquired with photon energies $\omega = 26.7$ cm$^{-1}$ (0.8 THz or 3.52 meV), corresponding to a wavelength λ = 375 µm. Gold pads contacting Hall bar devices were used as the optical reference material. AFM-SNOM scans with typical size of 1.3 µm × 8 µm (more than 2 orders of magnitude smaller than the photon wavelength) were taken in the temperature range 20 K < $T$ < 300 K. These images are shown in Fig. 2a,b. The near-field signal was demodulated up to the 4th harmonic of the AFM tapping frequency, see Supplementary Note 1. We focus here on the data corresponding to the 3rd harmonic, $S_3$.

In Fig. 2b we compare the temperature dependence of the near-field signal with that of the device resistance, $R(T)$. The corresponding temperatures are indicated to the right of each scan. LSCO regions are to the left and Au contact areas to the right in the panels. The dashed line between $T$ = 35 K and T = 30 K scans separates the data collected at temperatures collected above and below $T_c$. The error bars are primarily determined by surface roughness and the imperfections at the edge of the sample due to lithography. Good reproducibility in the data is demonstrated by scans acquired at the same temperatures on cooling and warming during different runs for the same contact and also at different LSCO-Au contacts. The main experimental observation is that, when probed with energies of $\omega$ = 26.7 cm$^{-1}$ (0.8 THz), the near-field intensity ratio between LSCO and Au changes substantially when the HTS sample enters the superconducting state, see Fig. 2c. Above $T_c$ this ratio is fairly temperature independent, staying below unity around a value ~0.95. There is an abrupt rise below $T_c$ where the relative intensity reaches a peak value close to 1.3 at $T \approx 25$ K. The signal decreases slightly but remains above unity upon further cooling to $T$ = 20 K. No such changes are observed if the sample is probed with mid-infrared photon energies $\omega \approx 1{,}000$ cm$^{-1}$ (30 THz or 125 meV), which are above the superconducting gap in LSCO.

**Theoretical modelling and interpretation**

The energy scale and the non-monotonic temperature dependence of the near-field signal across the superconducting transition provide strong constraints to possible interpretations of our data. Phonons do not display any abrupt changes at $T_c$. Moreover, the energies of optical and acoustic branches are either too high or too low to match the energy and range of momenta probed in our experiment[39]. Long wavelength and low energy magnetic excitations are already overdamped at much smaller Sr concentrations[40]. Local antiferromagnetic fluctuations persist at optimal doping[41] but they are at much higher energies, around 0.2 – 0.5 eV, and they also do not display any abrupt changes with temperature upon crossing $T_c$. Increased scattering due to larger far-field reflection coefficients in the superconducting state cannot account for the observations because of their temperature dependence, refs. [31,42], see also the discussion in Supplementary Note 5. We can also rule out electronic scenarios associated with both the critical behavior of the dielectric function around the percolation threshold[43,44]. A consideration of the energy scales, near-field coupling mechanisms, temperature dependence across the superconducting transition as well as of the parity and momentum selection rules can explain why scenarios invoking more exotic superconductivity-induced collective modes are very unlikely to account for our results[2,45-47] (a more detailed discussion in connection to these aspects can be found at the end of Supplementary Note 5).

In contrast, we show here that SJPWs explain observations quite naturally. This scenario accounts for all key features of our experiment: the coupling mechanism, the abrupt and non-monotonic behavior of the relative LSCO/Au signal with reversal of intensity upon crossing $T_c$ as well as the energy scale which is consistent with the Josephson plasma energy $\omega_J$ inferred from far-field measurements. Coupling to plasmonic excitations is enabled by evanescent modes waves with high in-plane momenta in the proximity of the AFM tip and it is well documented in the literature[48]. The peaks in the near-field signal are correlated to energies of surface modes. The dispersion $\omega(q)$ for the propagating electromagnetic mode confined to the planar interface between an isotropic and an optically uniaxial material is implicitly given by, see Supplementary Note 3:

$$q = \frac{\omega}{c}\sqrt{\frac{\varepsilon_1 \varepsilon_c(\omega)(\varepsilon_1 - \varepsilon_{ab}(\omega))}{\varepsilon_1^2 - \varepsilon_{ab}(\omega)\varepsilon_c(\omega)}} \qquad (2)$$

where $\varepsilon_1$ is the permittivity of the isotropic medium while $\varepsilon_{ab}(\omega)$ and $\varepsilon_c(\omega)$ are the *ab*-plane and *c*-axis dielectric constants of the anisotropic material. Note that if $\varepsilon_{ab} = \varepsilon_c$, we recover the textbook dispersion formula of surface polariton in isotropic media, see ref. [9] and Supplementary Equation (7). AFM-SNOM has thus access to both in- and inter-plane charge dynamics, even if just one surface is accessible. Note also that far-field techniques require either large single crystals with surfaces parallel to the *c*-axis or, possibly, a grazing-incidence configuration for the case of thin films where the *c*-axis is perpendicular to the surface. Because of deep-subwavelength resolution, crystals or flakes with µm-length dimensions are sufficient for AFM-SNOM.

The energies and spectral weights associated with Josephson plasma modes in our film can be seen in Fig. 3, which shows the energy-momentum plot of the Fresnel reflection coefficient $r_P(q,\omega)$. The AFM-SNOM momentum form factor is also shown in panel (a) of this figure by a white dashed line. Within the dipole model its functional form is given by $q^2 \cdot \exp[-2qz_0]$ where $z_0$ is the distance of minimum approach, chosen here to be equal to the tip radius $r_{tip} \sim 20$ nm, see also Supplementary Note 6. The closely spaced branches above unity represent propagating modes inside the film, while the feature at $\omega/\omega_J \sim 1$, nearly dispersionless in this momentum range, correspond to SJPWs. The vertical axis in Fig. 3a represents energy in units of $\omega_J = 56.7$ cm-1, which is the calculated Josephson plasma frequency for LSCO at $T = 0$ K. The values of $r_P(q,\omega)$ were calculated using measurements of *c*-axis reflectance in bulk crystals[31] and in-plane THz complex conductivity data in single-crystal LSCO films grown by ALL-MBE [34], see Supplementary Note 2. Our derivation of $r_P(q,\omega)$ generalizes the characteristic matrix formalism for treating anisotropic materials in stratified media, see Supplementary Note 4. This approach also helps generalize Eq. (2) for multilayers because even thin-film devices contain at least two interfaces: vacuum-LSCO and LSCO-substrate, in our case. In this configuration there are two SJPW branches whose energies at high momenta can be read off directly from the zeros of the denominator in Eq. (2), inserting for $\varepsilon_1$ the values corresponding to vacuum and the substrate, respectively, see also Supplementary Figure 3. A related approach has been used in ref. [49] for obtaining surface mode dispersions in topological semiconductors. An important point for our interpretation is the realization that, in spite of the very different physical origin, the near-field signal tracks with a good approximation the screened Josephson plasma frequency. The reason is that the asymptotic values of the SJPWs at high wave-vectors are pushed very close to $\omega_J$ due the strong anisotropy of the ab-plane and *c*-axis plasma frequencies in LSCO.

Thin films allow for the hybridization of surface modes that are either symmetric or anti-symmetric with respect to the reflection in horizontal symmetry plane. In the presence of a substrate these modes cease to have strict odd/even parity but crosstalk between them is in principle still allowed. In our experiment, i.e. for the range of momenta defined by our form-factor (see white dashed line in Fig. 3a) and energies right below the screened c-axis plasma, we are mostly sensitive to the surface mode confined to the top interface. This is because of the large and negative imaginary part of $k_{2z}$ in Supplementary Equation (10), the vertical component of the wave momentum inside the slab, which makes $r_P(q \approx 1/d_{film}, \omega \lesssim \omega_{P,c})$ practically equal to the vacuum-film reflection coefficient $r_{12}$. In other words, the pole in $\text{Im}[r_P(q,\omega)]$ associated with the mode propagating on the bottom interface has a vanishing strength. Further mathematical and visual explanations are given in Supplementary Notes 4 and 5 and Supplementary Figures 3 and 4.

Furthermore, the energy scale and the momentum distribution of our experiment (the white dashed line in Fig. 3a) also allow us to distinguish between bulk-cavity and surface modes. Propagating Josephson plasmons inside the film correspond to the higher energy branches in Fig. 3a while surface modes appear for our momentum range as a horizontal bright line. Fig. 3b shows that momentum integration across the full-width-at-half-maximum (FWHM) of the distribution does not affect the surface mode: the green and black dashed lines practically coincide at energies $\omega/\omega_J \sim 1$. However, it flattens out the peak structure of the bulk-cavity modes at higher energies. In conclusion, Fig. 3 reveals not only that modes propagating inside the film are at higher energy than the Josephson plasma edge, but also that momentum averaging completely smears out their spectral structure while leaving the surface modes intact because of their dispersionless nature at high momenta. We therefore find that our experiment is sensitive to SJPWs.

Last but not least, the scenario invoking surface Josephson plasmons is also able to explain the temperature dependence observed in Fig. 2b. The calculated temperature and frequency dependence of the near-field signal within the spherical dipole approximation is shown in Fig. 4, see also Supplementary Note 6. In this approximation the entire tip is replaced by a polarizable sphere whose radius is roughly given by the AFM tip apex curvature. The spherical-dipole approach underestimates the experimental contrast and has clear limitations because of its simplifying assumptions. While a quantitative agreement between the model and experiment is not anticipated, we nevertheless expect that the qualitative features of Fig. 2 are captured. Fig. 4 shows that this is indeed the case, attesting to the robustness of our interpretation in terms of SJPWs. The increase of the relative LSCO/Au signal above unity in the superconducting state as well as the non-monotonic temperature dependence are naturally understood in this scenario. As shown in Supplementary Figure 4, in our measurements we change the temperature and use the constant energy of our THz source. At a given temperature below $T_c$, the reflectance edge sweeps across our energy window and that is also the point where the near-field signal is enhanced. With further cooling, the superfluid density increases, the reflectance edge moves to higher energy, leading to a decrease in the near-field response, which is what we observe experimentally.

**DISCUSSION**

As expected, the spherical dipole model is not able to account quantitatively our data. For reasonable fitting parameters it renders values for the LSCO/Au intensity ratio that are too small. The shortcomings can be understood and are essentially related to the simplifying assumptions about effective tip polarizability, near-field distribution, field enhancement factors and the neglect of retardation effects. Indeed, increasing the tip polarizability or taking into account a more realistic shape of the AFM tip would lead to a larger LSCO/Au contrast than that shown in Fig. 4, see also Supplementary Figure 5 and the discussion in Supplementary Note 6. Substantial quantitative discrepancies related to the spherical dipole model were already remarked for experiments performed in the infrared range and more realistic models were proposed for the AFM tip geometry[50] as well as for the electrodynamic response of the sample-probe system[51,52]. Because in our experiment the wavelengths reach millimeter range, the shortcomings of this model are further exacerbated. Therefore, we believe that a meaningful quantitative agreement requires a full electrodynamical treatment taking accurately into account the shapes of AFM tips and the details of the scanning regime, which is outside the scope of this study.

Using the value of the Josephson plasma frequency $\omega_J = 0.8$ THz at $T = 0.8\ T_c$, at optimal doping we obtain a critical current density of the $c$-axis junction $J_c \sim 2\cdot10^5$ A/cm$^2$ and a Josephson penetration depth $\lambda_J = c/\omega_J \sim 375$ μm, indicating that our 20 μm wide wire is in a quasi-1D long Josephson junction regime. For insulating barriers, $\omega_J$ tracks, in turn, the superconducting gap $\Delta$. Using the normal-state conductivity evaluated from far-infrared reflectance data[31] the gap can be estimated[53] to $\Delta(T = 0.8T_c) \sim 24$ cm$^{-1}$ (0.75 THz), see also Supplementary Note 2. This is a value which is close to the frequency of AFM-SNOM data and it is in very good agreement with recent measurements of the superconducting gap in LSCO-based tunnel junctions[54]. Note that the momentum distribution probed in AFM-SNOM experiments depend on the probe geometry, see Fig. 3a. A systematic exploration of the strong coupling (polaritonic), regime of the SJPW modes is important for probing superconducting nodes or analysis of dissipation channels. This task can be accomplished in several ways. One is by taking advantage of the fact that parameters such as the conical shapes or apex radii can be engineered for etched wire tips[55,56]. Other routes can be provided by 'tilting' the light line dispersion by using thin cover layers of high dielectric permittivity such as SrTiO$_3$, or by using hyperspectral near-field imaging extended to the THz range[57]. It is important to note that Eq.(2), which deals with dispersion of excitations confined to planar interfaces, has to be amended when the wavelength of these modes is comparable to the sample width, which defines the Sommerfeld or Goubau regime[11,58,59]. A comparison with the momentum selective form-factor of the AFM tip, white dashed line in Fig.3, shows that in our configuration when the light couples to the sample only through the tip, the contribution of the surface modes in the long wavelength limit is vanishingly small. Consequently, Eq.(2) is applicable here for a broad momentum range which includes the relevant contributions for analysis. Note also that the extracted values for the Josephson penetration depth and superconducting gap are not obtained from direct measurements, but rather indirectly from the value of $\omega_J$ and theoretical considerations in ref. [53]. Measurements of samples with different widths combined with the use of engineered AFM tips with larger apex radii provide a clear path for checking experimentally the role of boundary conditions when exploring the strong light-matter regime.

The energy scales and the temperature dependence of the LSCO/Au near-field ratio constrain the interpretation sufficiently so that we can assign our observations to SJPWs. Other signatures compatible with this interpretation may exist in theory, for example real-space interference patterns of propagating plasmonic excitations[8]. We did not find observe such fringes experimentally. The reasons for their absence can be understood as the result of the $d$-wave gap and strong quasi-particle damping at temperatures $T \approx 0.8\cdot T_c$, which are close to the base temperature of our system. Customized AFM tip radii may be also needed to maximize the coupling efficiency. Plasma waves such as the ones reported here for LSCO have not been observed yet in Fe-based superconductors[60], although they are layered materials as well. It would therefore be interesting to explore the coupling to these excitations as a function of the sample anisotropy and damping in engineered heterostructures as well as in other cuprate families.

In addition to deep sub-diffractional resolution, light demodulation at different harmonics of the AFM tapping frequency provides dielectric depth information[21]. Thus, cryogenic AFM-SNOM allows probing both the lateral and vertical spatial (in)homogeneity in superconductors, offering a path for mapping superfluid profiles confined to buried interfaces[15]. A variable-temperature near-field environment can also enable characterization of confined modes in topological hetero-structures and quantification of length-scales associated to superconducting proximity effects in exposed surfaces or in buried interfaces[61]. Non-linear effects involving Josephson plasma in layered superconductors have been observed in LSCO by far-field methods[7,22] and they are predicted to be even stronger below $\omega_J$. It is conceivable that electromagnetic field enhancement around the tip will allow further avenues for exploration and control of such photonic effects in the THz range.

**Methods**

**A. $La_{1.85}Sr_{0.15}CuO_4$ (LSCO) film synthesis and device fabrication**.
Optimally-doped, LSCO films with a thickness of 13 nm were grown on LSAO substrates using ALL-MBE. The stoichiometry was controlled before and during deposition by a scanning quartz crystal monitor and by a calibrated custom-built 16-channel atomic absorption spectroscopy system, respectively. Film thickness was monitored in-situ by counting the reflection high-energy electron diffraction oscillations and post growth by AFM, profilometry and the finite-thickness oscillations in X-ray reflectance and diffraction. The latter also attest to the film being atomically smooth. The substrate temperature during the growth was 650 ºC and the ozone partial pressure was kept at $p = 2 \times 10^{-5}$ Torr. The device, part of which is shown in Fig. 1b was patterned using photolithography into a narrow strip with 32 pairs of lateral contact pads, as described in the main text. The data shown in Fig. 2 were acquired from an area in the vicinity of one such lateral contact (marked by a red rectangle in Fig. 1b) and the same results were reproduced for other contacts.

**B. AFM – SNOM measurements**.
Near-field and AFM data were acquired simultaneously in a custom-built AFM-SNOM setup. The sample chamber was optimized for laser light access to the AFM tip, sample visualization and a large numerical aperture in the collection path. AFM tips are reproducibly obtained by etching 100 μm diameter W wires in a 2M-NaOH solution and using a two-ring geometry. Voltages (peak-to-peak) in the 2 – 10 V range and frequencies between 5 Hz and 800 Hz can be used for shaping the AFM tip shaft and apex radius. The tip radius used here is ~20 nm as imaged by Scanning Electron Microscopy. The tips were glued to a piezo-actuated quartz tuning fork (TF) resonator operated at its resonance frequency of 28.38 kHz (the resonance frequency without the mounted tip was $2^{15}$ Hz = 32.768 kHz). AFM data were taken in amplitude-modulation mode with the feedback based on the TF piezocurrent. The resonance FWHM was ~ 5Hz and ~ 1Hz at room and low temperatures, respectively. Q-control was used to broaden the TF resonance when the system was cold. The AFM oscillation amplitude was $A \sim 80$ nm as determined from both AFM approach curves as well as from precise interferometric measurements. Light from a Backward Wave Oscillator (Microtech Instruments) with a wavelength $\lambda = 375$ μm (800 GHz = 26.7 cm$^{-1}$) was focused on the tip with the help of an off-axis Au-coated copper paraboloid. The SNOM signal was focused on a LN2 - LHe cooled hot electron bolometer and demodulated in real-time up to the 4th harmonic of the AFM tapping frequency with the help of a home-made data acquisition system. We analyzed the SNOM data within the dipole model, which roughly approximates the metallic AFM tip with a sphere whose polarizability is modulated by the interaction with the LSCO sample. Details about the model can be found in the Supplementary Note 6.

**C. AFM – SNOM Data Acquisition (DAQ).**
A custom made DAQ system was built in order to analyze the SNOM signal in-real time, simultaneously with AFM topography data. The hardware consists of a PC computer, a National Instruments high-speed digitizer and a microcontroller. Custom software provides the graphical user interface and synchronizes the operation of the system. The reference clock of the digitizer, reduced to its sampling frequency of a few MHz, is sent to the microcontroller which serves as a digital counter. The microcontroller also triggers the digitizer to start recording time-series data upon receiving pixel pulses from the ASC500 unit, which is the Attocube scanning probe microscope (SPM) controller. At each pulse, the microcontroller sends the current value of its counter to the PC to precisely identify each pixel within the digitizer's data streams. Two data streams are digitized and sent to the PC simultaneously: a reference sine wave output at the tuning fork's resonance frequency ($\Omega_{TF}$) from an output of the ASC500 controller and the actual detector signal. The reference has a dual purpose: (i) it allows for clock-independent measurements of $\Omega_{TF}$ for subsequent Fourier analysis of the detector signal at each pixel and (ii) it allows the SNOM-DAQ to operate the SPM in both amplitude and frequency modulation modes. A lock-in method allows demodulation of the SNOM signal at desired multiples of $\Omega_{TF}$. The calculated values for each pixel are then used to construct 2D dielectric SNOM images of the material surface.

## Data availability
The authors declare that all data supporting the findings of this study are available within the manuscript and the supplementary information. They are available from the corresponding author upon reasonable request.

## Code availability
All numerical codes in this paper are available upon request to the authors.

## Acknowledgements
A.G. and X.H. were supported by the Gordon and Betty Moore Foundation's EPiQS Initiative through Grant GBMF4410. A.T.B and I.B. were supported by the U.S. Department of Energy, Basic Energy Sciences, Materials Sciences and Engineering Division. A. G. also acknowledges support from the DOE Early Career Research program (Grant 2005410). Q.L and A.G. acknowledge support from the Yale West Campus Materials Characterization Core for SEM imaging of AFM tips.

## Author contributions
Q.L. and A.G. took and analyzed the SNOM-AFM data. The synthesis of LSCO films by ALL-MBE and characterization were done at Brookhaven by X.H., A.T.B and I.B.. Lithography and Hall bar characterization by transport was done by A.T.B. R.S. and A.G. designed and built the AFM-SNOM data acquisition system. A.G. led the experiment and wrote the manuscript with contribution from all authors.

**(*) Correspondence**: adrian.gozar@yale.edu

**Figure Legends**

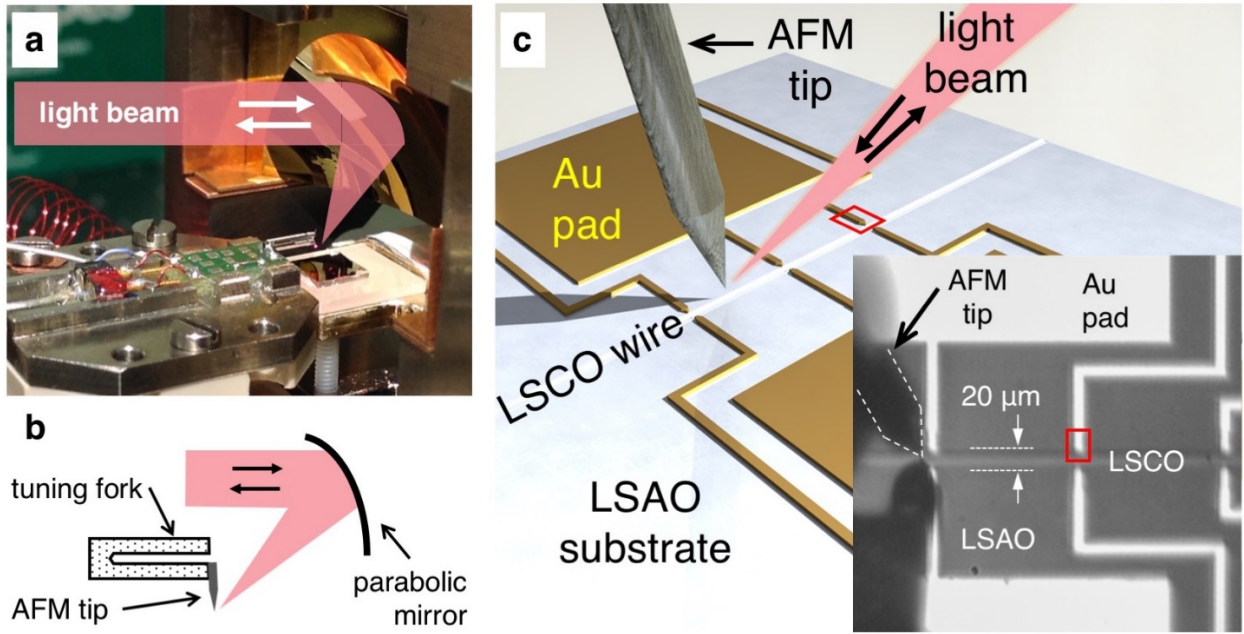

**Fig. 1 The experimental system and the measured device**. **a** A close-up of the cryogenic AFM-SNOM sample space showing the tuning fork mount, the sample holder and the path of the light beam reflecting off the parabolic mirror. The two arrows indicate light propagation directions in our back-scattering geometry. **b** Schematic showing the collimated beam being focused by the parabolic mirror onto the apex of the AFM tip. The AFM tip is an etched metallic wire glued to a prong of the quartz tuning fork sensor (see Methods). **c** The schematic of the measured device shows the light beam, the AFM tip, the Au pads, the 20 μm wide strip of 13 nm thick LSCO film and the LaSrAlO$_4$ substrate. During the measurements the tip was positioned next to lateral LSCO-Au contacts such as the one marked with a red rectangle. Similar data were obtained from all measured contact regions, attesting to the robustness of the results. The inset of this panel shows an actual CCD image of the investigated sample area. Visible on the left side of the inset is the etched AFM tip and its reflection in the substrate. The outline of the tip is emphasized by while dashed lines.

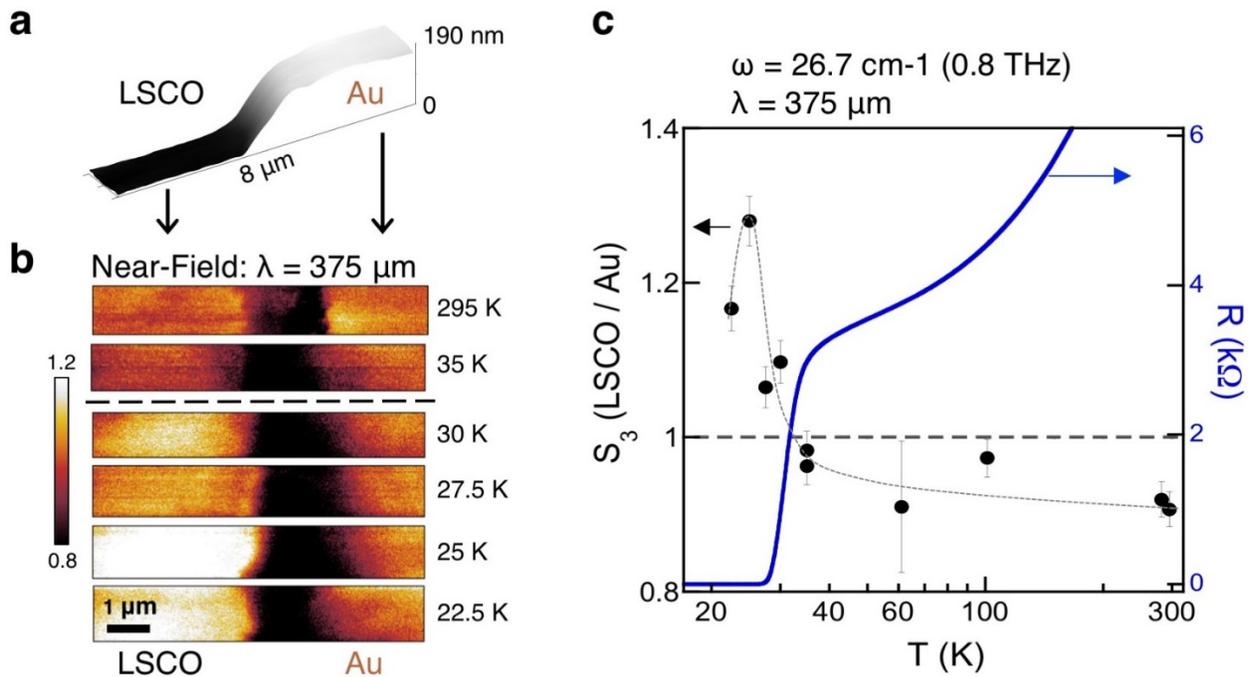

**Fig. 2 Temperature dependence of the near-field signal acquired with $\omega$ = 26.7 cm$^{-1}$ (0.8 THz) photon energy**. **a** 3D view of the AFM topography acquired at $T$ = 22.5 K around the LSCO-Au contact area. Dark/bright areas on the left/right correspond to LSCO and Au, respectively. **b** Temperature dependent near-field amplitude at several temperatures. LSCO and Au regions are indicated by arrows and labels at the bottom. The SNOM signal from the Au contact in each SNOM scan is normalized to unity. Dark areas between LSCO and Au are induced by the broad lithographic edge and were not used in the quantitative analysis. The scale bar in the bottom panel is 1 µm. **c** The black circles: near-field $S_3$ signal from LSCO normalized to Au, obtained from the data in panel (**b**) as a function of temperature, in log-scale. The grey-dashed line is a guide for the eye. The blue solid line: temperature-dependent resistance, $R(T)$, the right vertical scale.

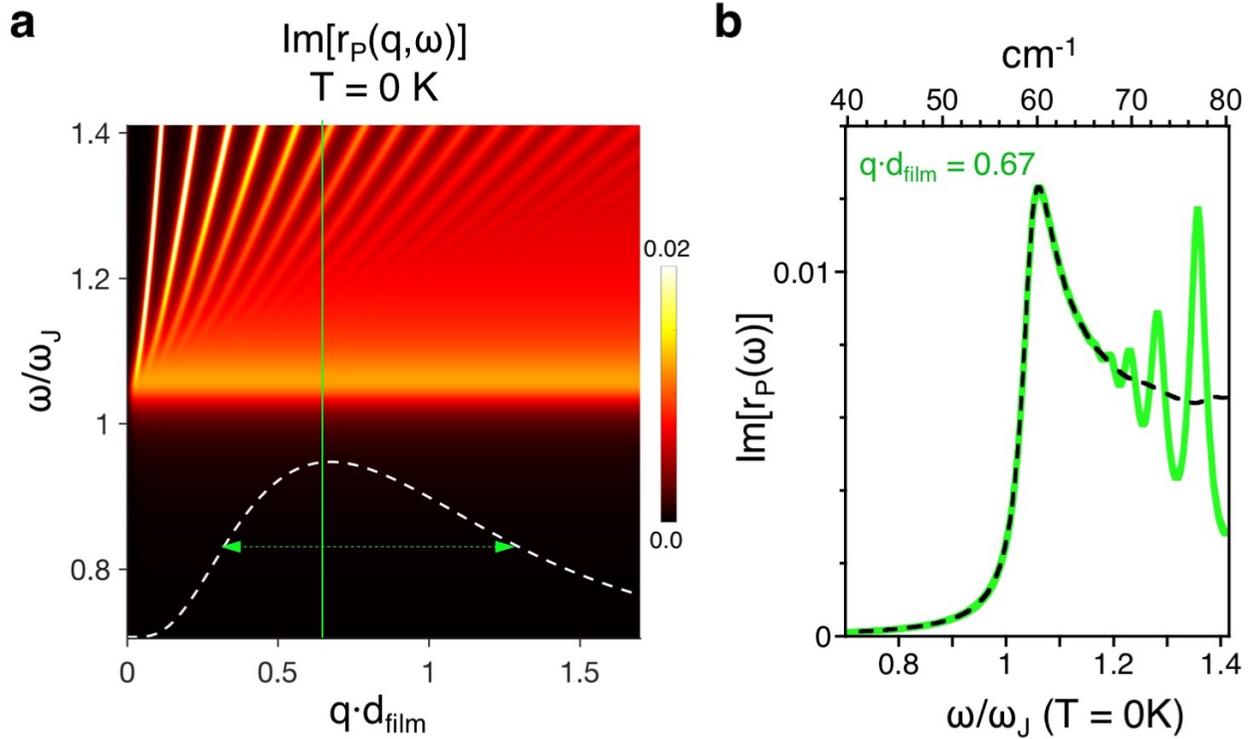

**Fig. 3 Dispersions of surface and guided Josephson plasmon modes**. **a** Contour plot of the calculated imaginary part of the Fresnel reflection coefficient $r_P(q,\omega)$ at zero temperature. The vertical axis is in units of $\omega_J$, the screened Josephson plasma frequency. The horizontal axis is in units of $q \cdot d_{film}$, where $q$ is the momentum and $d_{film}$ = 13 nm is the film thickness. The white dashed line is the AFM-SNOM momentum form-factor, see text and Supplementary Note 6. The green solid line is a constant-momentum cut at the maximum of the distribution (solid line) and the green dashed line marks the full-width at half-maximum (FWHM) of this distribution. **b** Constant momentum cuts of $\mathrm{Im}[r_P(q,\omega)]$ at $T$ = 0 K from the data in panel (a). The bottom axis is in dimensionless units $\omega/\omega_J$ and the top axis is in units of cm$^{-1}$. The green solid line corresponds to the momentum cut at the maximum of the form factor. The black dashed line is $\mathrm{Im}[r_P(\omega)]$ with integrated momentum across the FWHM of the form factor, as indicated by the green dashed line in (**a**).

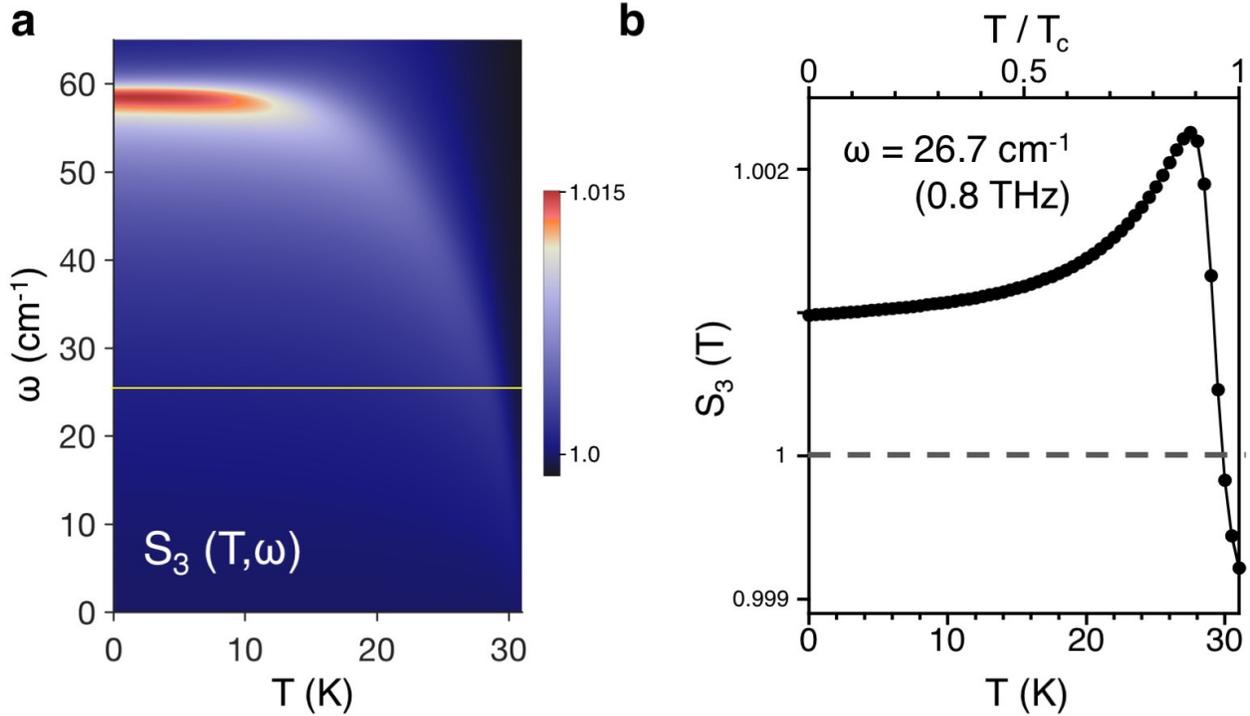

**Fig. 4 Temperature and frequency dependence of surface Josephson plasma waves**. **a** A contour plot of the calculated amplitude of the 3$^{rd}$ harmonic near-field signal from LSCO, $S_3(T,\omega)$, normalized to Au. The horizontal temperature scale extends from $T = 0$ K to the critical superconducting temperature $T_c = 31$ K. The yellow solid line marks the constant energy cut at $\omega = 26.7$ cm$^{-1}$ (0.8 THz). **b** Temperature dependence of $S_3(T)$ for $\omega = 26.7$ cm$^{-1}$, the yellow line in panel (**a**). This is the frequency at which the data in Fig. 2 were acquired. The grey dashed line marks equal near-field intensity from LSCO and Au, as in Fig. 2b. Our calculations reproduce well the experimental observations: a non-monotonic temperature dependence of the normalized $S_3(T,\omega)$ signal and the reversal of contrast in SNOM signal between LSCO and Au upon entering the superconducting state. This is due to the resonant excitation of SJPWs in the LSCO film.

**Supplementary Information for**

# Surface Josephson plasma waves in a high-temperature superconductor

Q. Lu[1,2,3], A.T. Bollinger[4], X. He[4,5], R. Sundling[4,6], I. Bozovic[2,4,5] and A. Gozar[1,2,7,*]

[1] Department of Applied Physics, Yale University, New Haven, CT 06520, USA
[2] Energy Sciences Institute, Yale University, West Haven, CT 06516, USA
[3] Shaanxi Institute of Flexible Electronics, Northwestern Polytechnical University, Xi'an, 710072, China
[4] Brookhaven National Laboratory, Upton, NY 11973, USA
[5] Department of Chemistry, Yale University, New Haven, CT 06520, USA
[6] Zensoft, Inc., Madison, Wisconsin 53705, USA
[7] Department of Physics, Yale University, New Haven, CT 06520, USA

* adrian.gozar@yale.edu

## Supplementary Note 1.
## Additional information about AFM-SNOM data: near-field interaction and spatial resolution of our measurements

Supplementary Figure 1 demonstrates that our optical signal originates in the true near-field interaction between the AFM tip and the sample. The near-field interaction is non-linear as a function of the tip-sample distance[1,2,3-5]. As a result, the reflected signal obtained when the AFM is operated in tapping mode contains harmonics of the AFM resonance. Supplementary Figure 1a shows a typical AFM-SNOM sample approach signal demodulated up to the 4$^{th}$ harmonic of this frequency. All harmonics display very good suppression of the background component, attesting thus for the true near-field character of our optical measurements. Panel (b) in Supplementary Figure 1 shows the ratio of the near-field signal corresponding to the 3$^{rd}$ and 2$^{nd}$ harmonics while the AFM tip is in intermittent contact to the sample. This ratio is around 0.25 for both Au and LSCO and it is only very weakly temperature dependent. This value can be used in principle for refining models for the tip-sample interaction, e.g. to help fix adjustable parameters in various theoretical approaches[4,6].

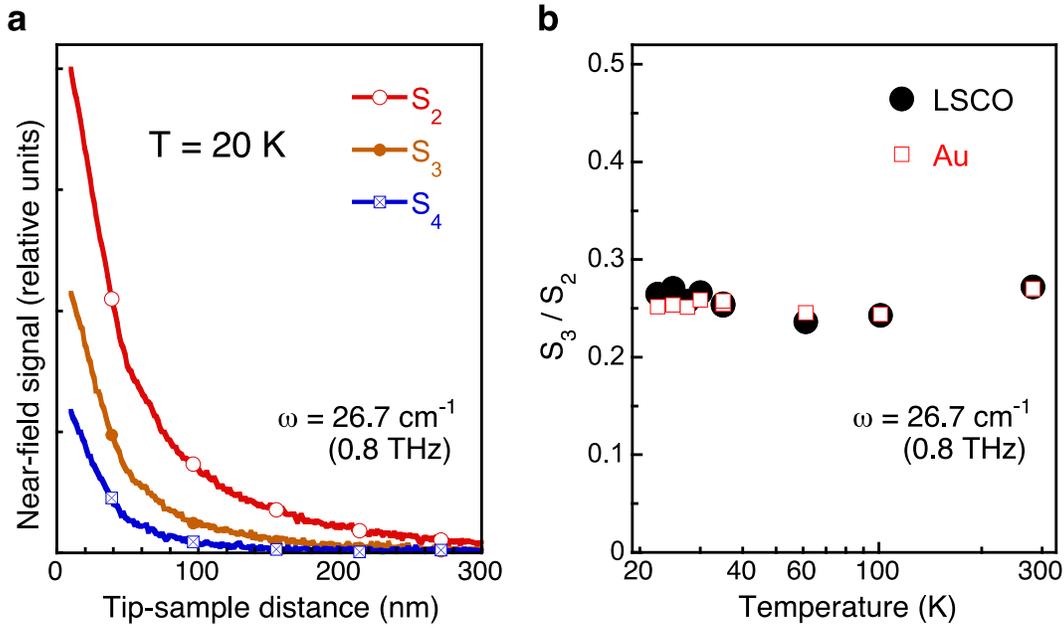

**Supplementary Figure 1. Near-field signal at several harmonics of the AFM resonance frequency.** **a** Low-temperature ($T$ = 20 K) approach curves for Au contacts with the light energy $\omega$ = 26.7 cm$^{-1}$ (0.8 THz). $S_2$, $S_3$ and $S_4$ are signals demodulated at the 2$^{nd}$, 3$^{rd}$ and 4$^{th}$ harmonics of the AFM tapping frequency of 28.02 kHz. **b** Relative intensities of the maxima of the 2$^{nd}$ and 3$^{rd}$ harmonics for LSCO (solid black circles) and Au (open red squares) as a function of temperature (shown in logarithmic scale) when the AFM tip is in intermittent contact to the sample.

Supplementary Figure 2 provides information about the spatial resolution of our measurements. The left panel shows AFM topography and near-field data in a single layer graphene sitting on top of a SiO$_2$ substrate. The line profile shown in panel (b) of this figure shows a transition region of about 100 – 150 nm. Given that our probing wavelength is $\lambda$ = 375 μm we obtain a resolution round $\lambda$/2500 – $\lambda$/3750 consistent with our statement in the main text. The data shown in Supplementary Figure 2 were taken with our typical tips having apex radii R ≈ 20 nm. No particular care was taken to obtain either extra sharp (R ≤ 5 nm) tips.

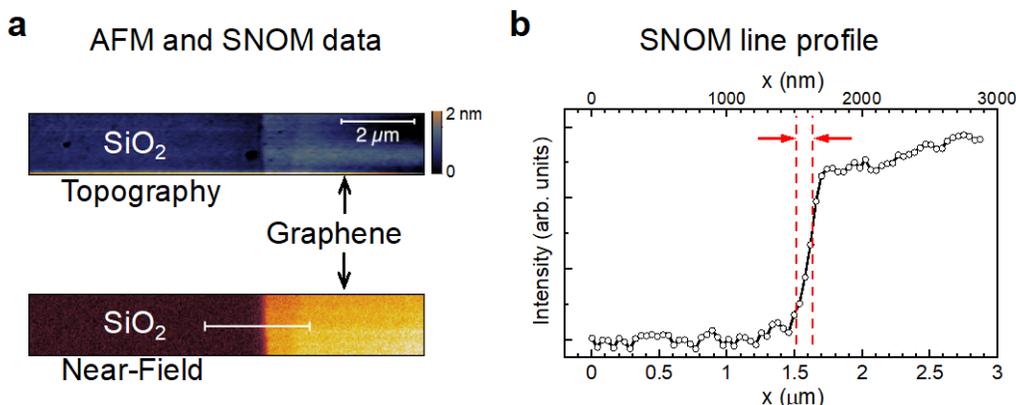

**Supplementary Figure 2. Spatial resolution a** AFM topography (top) and near-field data (bottom) at the edge of a single layer graphene flake on a SiO2 substrate. Graphene appears brighter than the oxide. **b** Near-field line profile along the line shown in the scan on the bottom of panel (b). The vertical red dashed lines indicate 10% and 90% of the transition region and were used to determine the spatial resolution.

## Supplementary Note 2.
## Anisotropic optical data analysis in La$_{1.85}$Sr$_{0.15}$CuO$_4$

The calculated dispersions of the Josephson Surface Plasma Waves (JSPWs) and of the AFM-SNOM response in our LSCO film, see Figures 3 and 4 in the main text, require knowledge of the dielectric constants in directions parallel and perpendicular to the CuO$_2$ planes. For $c$-axis data, we used the results in refs. [7] and refs. [8-10]. For the $ab$-plane, we used the results obtained by time-domain THz spectroscopy[11] correlated to reflectivity measurements in LSCO films with the same doping[12,13]. The real part of the conductivity $\sigma_1(\omega)$ was fitted with a Lorentzian (Drude model) at selected temperatures. The superfluid weight was obtained by fitting the imaginary part of the conductivity $\sigma_2(\omega)$ to the function $F(\omega) = A/\omega + \sigma_{dc}\Gamma\omega/(\omega^2 + \Gamma^2)$, where the dc conductivity $\sigma_{dc}$ and the scattering rate $\Gamma$ were determined from the fits of the real part. The superfluid plasma frequency is $\omega_{pS,ab} \propto \sqrt{A}$. The data were then interpolated at all temperatures with suitable polynomials. The values $\varepsilon_{\infty,ab} = 4$ and $\varepsilon_{\infty,c} = 25$ were chosen based on refs. [7,8,12,13].

For the Au reference, we used the data from refs. [14] and [15]. The value $\varepsilon_{Au}$ = -738752 + i·2077715 was chosen based on the plasma frequency $\omega_{p,Au}$ = 8.5 eV and the scattering rate $\Gamma_{Au} \approx 75$ cm$^{-1}$, for $T \approx 77$ K. At room temperature, $\Gamma_{Au}$ is a factor of ≈ 5 larger, as indicated by transport measurements. Reasonably large variations in the real and imaginary parts of $\varepsilon_{Au}$ were found to have no impact on the overall results. The dielectric function for the etched Tungsten tip[16] and LSAO substrate[17] used for the dielectric function at $\omega \sim 26.7$ cm$^{-1}$ are: $\varepsilon_W$ = -12377 + i·12319 and $\varepsilon_{LSAO}$ = 19.

These optical parameters also allowed us to estimate the $c$-axis normal state resistivity. Together with the value of the screened Josephson plasma edge $\omega_J$, which allows determination of the critical current along this axis, one could try to estimate the value of the superconducting gap $\Delta$ using the Ambegaokar-Baratoff formula, see the discussion section in the main text and ref. [18]. The set of equations we used are:

$$\omega_P = \sqrt{\frac{2\pi I_c}{\Phi_0 C}} \quad (1a) \qquad \sigma = \frac{\omega_P^2}{60 \cdot \Gamma} \quad (1b) \qquad I_c R_n = \frac{j_c d}{\sigma} \quad (1c) \qquad I_c R_n = \frac{\pi \Delta}{2e} \tanh\left[\frac{\Delta}{2k_B T}\right] \quad (1d)$$

The critical current density $j_c$ was estimated from Supplementary Equation (1a) using $I_c = j_c \cdot S$ and $C = \varepsilon_r \varepsilon_0 S/d$, where $C$ is the capacitance between two CuO$_2$ planes and d = 6.62 Å is the distance between two CuO$_2$ planes. The relative permittivity $\varepsilon_r$ was determined from the far-IR reflectance data in ref. [7]. The normal carrier conductivity $\sigma$ was obtained from Supplementary Equation (1b) which is the DC limit of the Drude response, see Eq. (1) in the main text. Supplementary Equation (1b) above gives $\sigma$ in units of $\Omega^{-1}\cdot$cm$^{-1}$ when $\omega_P$ and $\Gamma$ are expressed in units of cm$^{-1}$ (here $\omega_P \approx$ 300 cm$^{-1}$ and $\Gamma \approx$ 250 cm$^{-1}$ from ref. [7]). Using Supplementary Equation (1a) and (1b) we determined the $I_c \cdot R_n$ product from Supplementary Equation (1c) and then the superconducting gap from (1d).

**Supplementary Note 3.**
**Dispersion of surface waves at the interface between an isotropic and uniaxial material**

Here we derive Eq. (2) in the main text, which is a generalization of the textbook relation for surface plasmons, see for example ref. [19]. The geometry of the problem is similar to that of Supplementary Figure 3a, but with only the top interface present. Maxwell's equations for time harmonic fields read:

$$\nabla(\hat{\varepsilon}\mathbf{E}) = 0 \,, \quad \nabla \mathbf{B} = 0 \,, \quad \nabla \times \mathbf{E} = i\omega \mathbf{B} \quad \text{and} \quad \nabla \times \mathbf{B} = -i\omega \hat{\varepsilon} \mathbf{E}/c^2 \quad (2)$$

where the free carrier contribution is included in the complex dielectric constant shown in Equation (1) of the main text. The currents and surface charges can be obtained from $\mathbf{j} = \sigma \cdot \mathbf{E}$ and $\sigma_{\text{surf}}(x) = \varepsilon_0 \cdot [E_{1z}(x) - E_{2z}(x)]$ where the conductivity σ is a tensor with components given by $\varepsilon_\alpha = \varepsilon_{\infty,\alpha} + i\sigma_\alpha/\varepsilon_0 \omega$ where α is the ab-plane or c-axis index. It also has two components corresponding to the normal and superfluid parts, respectively. The equations and boundary conditions allow for separation of the transverse electric and magnetic modes. We confine ourselves to the latter as surface modes can be excited only with p-polarized light, i.e. $E_y \equiv B_x \equiv B_z = 0$. We look for propagating solutions of Maxwell equations, with the electric and magnetic fields decaying exponentially away from the interface:

$$\mathbf{E}_j(x,z) = (E_{jx}\hat{i} + E_{jz}\hat{k}) \cdot e^{iqx} e^{-k_j z} \,, \quad j = 1, 2 \text{ with } k_j \in \mathbb{R} \,, k_1 > 0 \text{ and } k_2 < 0 \quad (3)$$

where '$j$' is an index for the two media. Gauss' laws and the boundary conditions give:

$$iq E_{1x} = k_1 E_{1z} \,, \quad iq \varepsilon_{ab} E_{2x} = k_2 \varepsilon_c E_{2z} \,, \quad E_{1x} = E_{2x} \quad \text{and} \quad \varepsilon_1 E_{1z} = \varepsilon_c E_{2z} \quad (4)$$

Supplementary Equation (4) is a set of four homogeneous equations with four unknowns: $E_{1x}$, $E_{1z}$, $E_{2x}$ and $E_{2z}$. The condition for non-trivial solutions and the bulk dispersion relations for the isotropic and uniaxial medium read:

$$\frac{k_1(q,\omega)}{k_2(q,\omega)} = \frac{\varepsilon_1(\omega)}{\varepsilon_{ab}(\omega)} \text{ , where } q^2 - k_1^2 = \frac{\omega^2}{c^2}\varepsilon_1 \text{ and } \frac{q^2}{\varepsilon_c} - \frac{k_2^2}{\varepsilon_{ab}} = \frac{\omega^2}{c^2} \quad (5)$$

Supplementary Equation (5) is used for deriving the dispersion relations for the surface modes. The first equality in this equation is equivalent to:

$$q = \frac{\omega}{c}\sqrt{\frac{\varepsilon_1 \varepsilon_c(\omega)[\varepsilon_1 - \varepsilon_{ab}(\omega)]}{\varepsilon_1^2 - \varepsilon_{ab}(\omega)\varepsilon_c(\omega)}} \quad (6)$$

which is Equation (2) of the main text. If both materials are isotropic, i.e. $\varepsilon_{ab} = \varepsilon_c = \varepsilon_2$ one obtains the standard textbook formula for the dispersion of the surface plasmons:

$$q = \frac{\omega}{c}\sqrt{\frac{\varepsilon_1 \varepsilon_2(\omega)}{\varepsilon_1 + \varepsilon_2(\omega)}} \quad (7)$$

The asymptotic value of the dispersion $\omega(q\to\infty)$ in Supplementary Equation (6), i.e. the electrostatic limit, is given by the condition of vanishing denominator $\varepsilon_1^2 - \varepsilon_{ab}(\omega)\cdot\varepsilon_c(\omega) = 0$. The same equation can be used for the bottom interface in Supplementary Figure 3a with $\varepsilon_1$ replaced by $\varepsilon_3$. They determine the positions of the green dashed lines in Supplementary Figure 2b-d. Neglecting the dissipative contribution part, for a metal one can write $\varepsilon_{ab} \approx \varepsilon_{\infty,ab} - \omega_{P,ab}^2/\omega^2$ and $\varepsilon_c \approx \varepsilon_{\infty,c} - \omega_{P,c}^2/\omega^2$. If the anisotropy is large ($\omega_{P,ab} / \omega_{P,c} \gg 1$, a condition well satisfied for cuprates) a zero of the denominator in Supplementary Equation (6), i.e. the asymptotic energy of the surface mode for high momenta, is given with a very good approximation by $\omega_{P,c}/\sqrt{\varepsilon_{\infty,c}}$. This is marks the energy of the reflectivity edge, which in the superconducting state is the Josephson plasma energy $\omega_J$. This is why the green dashed lines in the anisotropic case of Supplementary Figure 3c are pushed closer to $\omega_J$ compared to the plots in panels (a) and (b) of the same figure. This is also the reason why the near-field signal has a peak which is very close to the reflectivity edge, see Supplementary Figure 4.

## Supplementary Note 4.
## Dispersions of bulk and surface plasma waves in films

The JPWs are non-radiative modes, i.e. for these excitations the electromagnetic fields are exponentially damped outside of the film. They are also true normal modes of the system. Here we provide derivations for the basic formulas in connection to guided and surface JPWs in films, see Fig. 3 in the main text. They appear in various forms in previous publications[20-31]. Typically they are obtained by writing down the field solutions along with the corresponding boundary conditions and by looking for non-trivial solutions of the system of homogeneous equations, i.e. a similar approach as the one in Supplementary Equations (3) and (4).

In this work we generalize the characteristic matrix formalism[32] for extraction of the reflection coefficient $r_P(q,\omega)$ to the case of anisotropic media. This approach is useful because it can be applied easily to larger number of layers, which corresponds to the typical experimental situation. In this section we also illustrate qualitatively the evolution of the dispersions of the two surfaces modes as a function the dielectric properties of the surrounding media. The characteristic matrix for an anisotropic medium such as the film in Supplementary Figure 3a is given by:

$$\begin{bmatrix} E_x(x,0) \\ E_z(x,0) \end{bmatrix} = \begin{bmatrix} \cos(k_z z) & ip\sin(k_z z) \\ \frac{i}{p}\sin(k_z z) & \cos(k_z z) \end{bmatrix} \begin{bmatrix} E_x(x,z) \\ E_z(x,z) \end{bmatrix} \quad \text{where} \quad p = \frac{\varepsilon_c k_{2z}}{\varepsilon_{ab} q} \quad (8)$$

The transmitted / reflected fields are at $z = 0$ and $z = d$, respectively. Using the boundary conditions along with Fresnel equations, Supplementary Equation (8) becomes:

$$\begin{bmatrix} -\frac{k_{3z}}{\sqrt{\varepsilon_3}} t \\ \frac{q\sqrt{\varepsilon_3}}{\varepsilon_c} t \end{bmatrix} = \begin{bmatrix} \cos(k_z d) & ip\sin(k_z d) \\ \frac{i}{p}\sin(k_z d) & \cos(k_z d) \end{bmatrix} \begin{bmatrix} -\frac{k_{1z}}{\sqrt{\varepsilon_1}}(1-r) \\ \frac{q\sqrt{\varepsilon_1}}{\varepsilon_c}(1+r) \end{bmatrix} \quad (9)$$

where $r = r_P(q,\omega)$ is the reflection coefficient plotted in Fig. 3 of the manuscript. It is given by:

$$r_P(q,\omega) = \frac{r_{12} + r_{23}e^{-2ik_{2z}d}}{1 + r_{12}r_{23}e^{-2ik_{2z}d}} \quad \text{with} \quad r_{12} = \frac{\varepsilon_{ab}k_{1z} - \varepsilon_1 k_{2z}}{\varepsilon_{ab}k_{1z} + \varepsilon_1 k_{2z}} \;,\; r_{23} = \frac{\varepsilon_3 k_{2z} - \varepsilon_{ab}k_{3z}}{\varepsilon_3 k_{2z} + \varepsilon_{ab}k_{3z}} \quad (10)$$

and

$$k_{1z} = -\sqrt{\frac{\omega^2}{c^2}\varepsilon_1 - q^2} \;,\; k_{2z} = -\sqrt{\varepsilon_{ab}\frac{\omega^2}{c^2} - \frac{\varepsilon_{ab}}{\varepsilon_c}q^2} \quad \text{and} \quad k_{3z} = -\sqrt{\frac{\omega^2}{c^2}\varepsilon_3 - q^2} \quad (11)$$

The '−' signs here only account for the downward propagation of light, opposite to the positive direction of the $z$-axis as defined in Supplementary Figure 3a. The poles of $r_P(q,\omega)$ give the implicit equation for the dispersion of SJPWs:

$$\tan[k_2 d] = \frac{i\varepsilon_{ab}k_{2z}[\varepsilon_3 k_{1z} + \varepsilon_1 k_{3z}]}{\varepsilon_{ab}^2 k_{1z}k_{3z} + \varepsilon_1 \varepsilon_3 k_{2z}^2} \quad (12)$$

An illustration of the meaning of Supplementary Equations (8 –12) is given in Supplementary Figure 3. The basic trends of these modes as a function of the dielectric environment are shown qualitatively in panels (b), (c) and (d) of this figure. For symmetric configurations ($\varepsilon_1 = \varepsilon_3$) the modes can be classified by the parity of the fields with respect to a reflection in a plane parallel to the surface at $z = d/2$. Supplementary Equation (12) separates into two equations, one for symmetric modes with respect to the '$x$' component of the electric field, i.e. $E_x(x, z + d/2) = E_x(x, -z + d/2)$, and the other for antisymmetric modes. The symmetric / antisymmetric modes correspond to the lower / upper branches, respectively, in Supplementary Figure 3b. The asymptotic value $\omega(q \to \infty)$ for the symmetric case, green dashed line in Supplementary Figure 3b, is given by the vanishing denominator condition in Supplementary Equation (6): $\varepsilon_1^2 - \varepsilon_{ab}(\omega) \cdot \varepsilon_c(\omega) = 0$. If $\varepsilon_{ab} \neq \varepsilon_c$, the symmetry is lost, and the asymptotic values are given by $\varepsilon_k^2 - \varepsilon_{ab}(\omega)\cdot\varepsilon_c(\omega) = 0$ with $k = 1,3$. The top / bottom branches in Supplementary Figure 3c correspond in the $q\cdot d_{film} \to \infty$ limit to non-interacting surface modes residing at the top / bottom interfaces respectively.

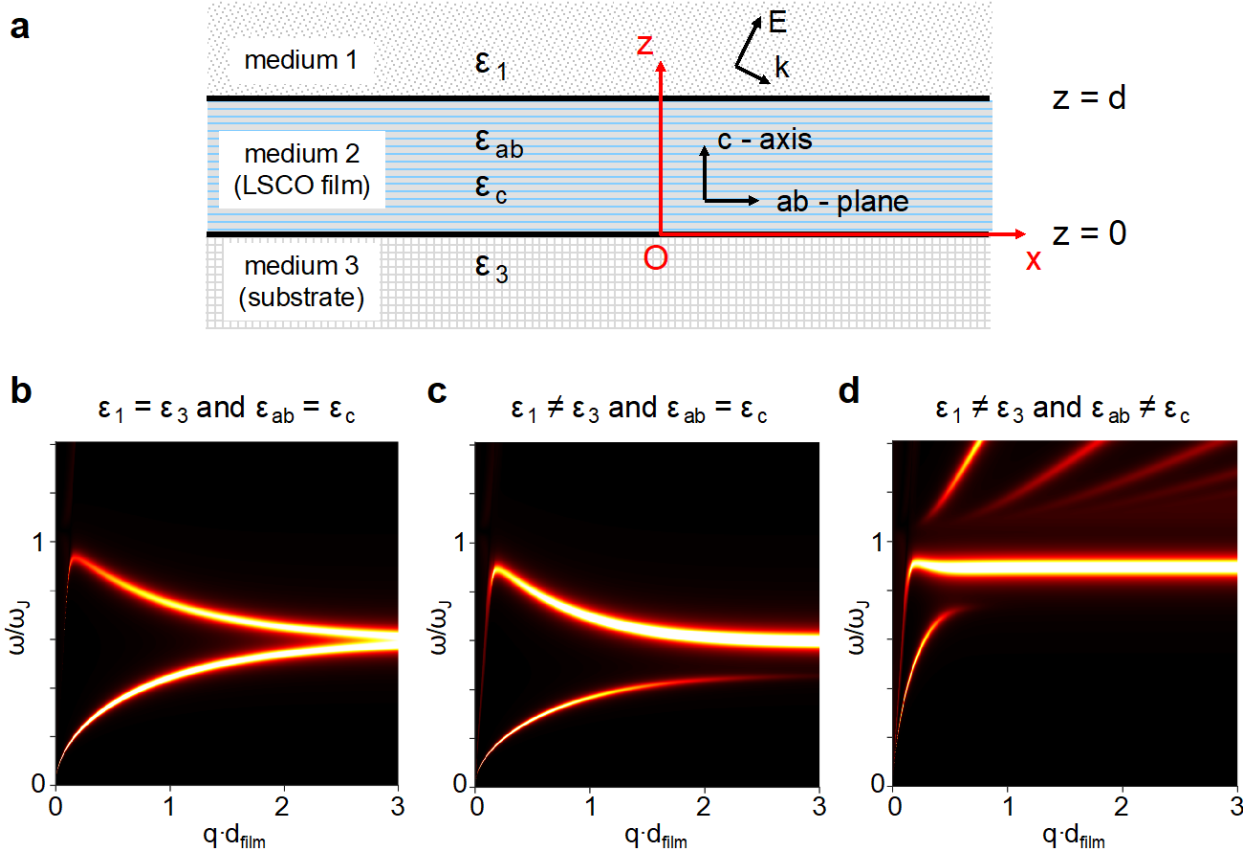

**Supplementary Figure 3. The effect of substrate and of the electronic anisotropy on the dispersion of Josephson plasma waves**. **a** Schematic of the experimental three-layer structure. Panels **b – d** show the results of calculations of Im[$r_P(q,\omega)$] at $T = 0$ K, illustrating qualitatively the evolution of the dispersions of surface JPWs from the isotropic ($\varepsilon_{ab} = \varepsilon_c$) and symmetric ($\varepsilon_1 = \varepsilon_3$) configuration to an anisotropic ($\varepsilon_{ab} \ne \varepsilon_c$) and non-symmetric ($\varepsilon_1 \ne \varepsilon_3$) arrangement. The following parameters were chosen for better visualization: $\varepsilon_1 = \varepsilon_3 = 50$ and $\varepsilon_{ab} = \varepsilon_c = \varepsilon_{c,LSCO}$ ($T = 0$ K) for panel (b); $\varepsilon_1 = 50$, $\varepsilon_3 = 100$ and $\varepsilon_{ab} = \varepsilon_c = \varepsilon_{c, LSCO}$ ($T = 0$ K) for panel (c); $\varepsilon_1 = 50$, $\varepsilon_3 = 100$ and the ab-plane superfluid plasma frequency increased to three times the value along the c-axis, $\omega_{pS,ab} = 3\cdot\omega_{pS,c}$ for panel (d). The energy is in units of the screened Josephson plasma frequency $\omega_J$. The film thickness $d = 500$ nm was chosen for all panels. Green dashed lines in each panel correspond to the asymptotic values of the SJP energies in the large $q^*d_{film}$ limit. The tilted grey dashed line is the light dispersion, $\omega(q) = q\cdot c/\sqrt{\varepsilon_1}$.

As discussed at the end of Supplementary Note 3, anisotropy in the plasma frequencies pushes the asymptotic energies of both these branches close to $\omega_J$. In Supplementary Figure 3d, a relatively small ratio $\omega_{p,ab} / \omega_{p,c} = 3$ was chosen, but the effect is nevertheless substantial. For LSCO this ratio is ~ 25, almost one order of magnitude higher, and this is why the energies probed in our experiment are so close to $\omega_J$, see Fig. 3 in the main text and Supplementary Figure 4. Inter-layer electron tunneling gaps the dispersions of the propagating modes in a superlattice[33], which otherwise obey $\omega(q\to 0, k_z \neq 0) \to 0$. Furthermore, these guided modes appear only when anisotropy is present, see Supplementary Figure 3d, because propagation inside the film along the z-axis is only possible for frequencies where the ab-plane and c-axis dielectric constants have opposite signs, which is the so-called hyperbolic regime. This can be seen by inspection of the second equality in Supplementary Equation (11).

**Supplementary Note 5.**
**Additional information about the interpretation in terms of Surface Josephson Plasma Waves (SJPWs)**

The interpretation of our data in terms of SJPWs rests on three pillars: (1) the intimate connection to superconductivity: the changes in the near-field signal are activated by the material entering the superconducting state; (2) the energy scales and (3) the non-monotonic temperature dependence of the relative near-field signal of LSCO compared to Au. The LSCO/Au ratio goes above unity upon crossing $T_c$ and then it decreases again displaying a non-monotonic behavior. We found that these observations provide sufficient constraints for the interpretation. In the main text we mentioned other scenarios that could be, in principle, invoked to explain our results. In this section we look at these aspects in more detail. We start by showing why surface Josephson plasma waves can explain the experimental findings in a straightforward way. We then discuss and rule out the possible role played by far-field reflection coefficients, inhomogeneous behavior close to $T_c$ as well as scenarios involving more exotic superconductivity-induced collective modes

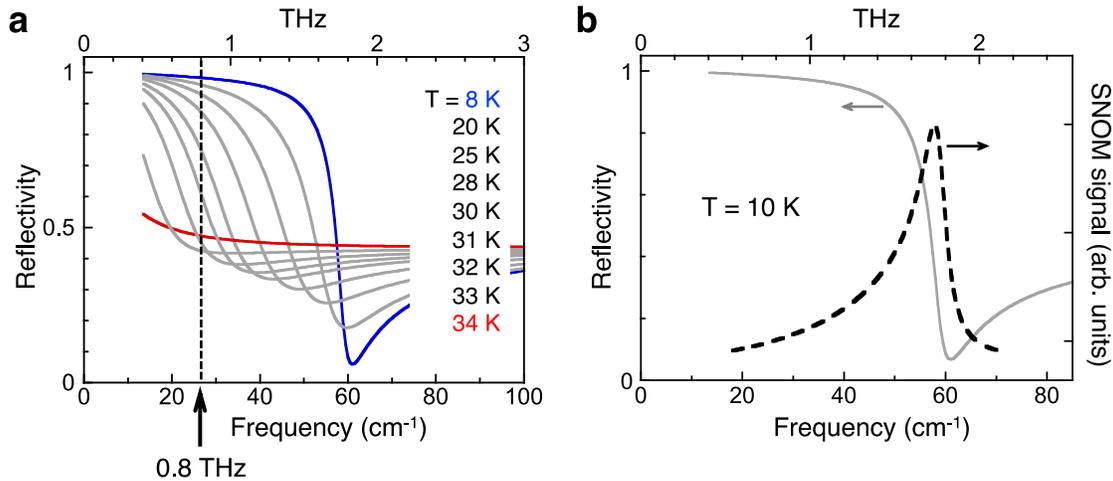

**Supplementary Figure 4. The connection between the plasma edge seen in the far-field reflectance data and the resonant enhancement of the near-field signal. a** Reflectance data with electric field polarized along the c-axis at various temperatures in optimally doped LSCO. The grey lines are obtained from the fitting parameters provided in ref. [7]. The vertical dashed line corresponds to our measurement frequency $\omega = 0.8$ THz $= 26.7$ cm$^{-1}$. **b** Reflectivity at T = 10 K from panel (**a**) along with the calculated near-field signal at the same temperature (black dashed line).

In the interpretation section of the main text we also argued that SJPWs provide a natural explanation for the data in Fig. 2 of the main text. Here we explain why this is so. The basis for our discussion is Supplementary Figure 4. Panel 4b of this figure shows that the near field signal at a given temperature is peaked very close to the reflectivity edge. In that sense one can say that SNOM signal tracks approximately the frequency of the plasma edge. However, in spite of similar energies it is crucial to realize that the origin of these two signals is very different. The SNOM signal is peaked at energies corresponding to the poles of the Fresnel reflection coefficient $r_P(q,\omega)$, i.e. at energies of surface modes. The reflectance edge is given by energy of the screened plasma frequency, $\omega_J = \omega_P/\sqrt{\varepsilon_\infty}$, which is a very different quantity. The reason these two values are so close to each other is the very anisotropic nature of the sample ($\omega_{P,ab}/\omega_{P,c} \approx 25$ at low temperatures for LSCO) and it is explained mathematically at the end of Supplementary Note 3 and graphically in Supplementary Figure 3d. The idea is that the anisotropy makes the asymptotic value of surface modes energy at high momenta, a pole of $r_P(q,\omega)$, to be very close to the screened c-axis plasma frequency.

With this discussion we can see how Supplementary Figure 4a explains the temperature dependence of the near-field signal in Fig. 2 of the main text. In our measurements we change the temperature and use the constant frequency of our THz source, the black dashed line in panel (a). At some temperature below $T_c$ the reflectance edge sweeps across our energy window and that is the point where the SNOM signal is enhanced. As the superfluid plasma frequency increases further with cooling, the reflectance edge moves to higher energies and the SNOM signal decreases again, showing thus non-monotonic behavior.

Increased scattering due to larger far-field reflection coefficients in the superconducting state cannot account for the observations. Similarly to a normal metal, the temperature-induced changes of the far-infrared in-plane reflectivity in optimally doped LSCO are very small: by less than 2% on cooling from room temperature to $T = 5$ K and by less than 1% below $T_c$, ref. [12]. Such changes are very gradual and also barely detectable by conventional techniques, so they cannot be responsible for the observed effect. More dramatic changes occur along the c-axis due to Josephson tunneling[7]. However, an explanation in terms of increased far-field c-axis reflectance is incompatible with the non-monotonic temperature dependence of the near-field signal below $T_c$. Lowering the temperature will only increase the brightness of the sample which becomes monotonically more reflective with cooling. Furthermore, because our wavelength ($\lambda = 375$ μm) is much larger than our scanning areas it is difficult to understand how far-field reflections from a relatively large diffraction limited spot can produce effects that will not divide out in the LSCO/Au ratio as we move the AFM tip only slightly from one material to another. In fact, the approach curves shown in Supplementary Figure 1 clearly demonstrate that the sample is probed in the near-field, which also means a distribution of finite momenta for the excitations as shown in Fig.3a of the main text. In conclusion, far-field effects can be ruled out.

We conclude this Supplementary Note by discussing other electronic scenarios compatible, in principle, with near-field coupling. Associating our results with the critical behavior of the dielectric function around the percolation threshold[34] is not a viable option. The fluctuation peaks seen in the 10's of GHz range[35] do not survive in the THz region[11]. Consistent with these expectations, in spite of our deep sub-diffraction capabilities, see Supplementary Figure 2, we did not detect experimentally inhomogeneous behavior in our optimally doped sample. The only other conceivable candidates in the electronic channel that are activated by the superconducting transition are those associated either with the c-axis Josephson tunneling, discussed in the next paragraph, or with the opening of the d-wave gap and the appearance of superconductivity-induced collective modes. Examples are Carlson-Goldman excitations[36], amplitude oscillations of the order parameter (Higgs modes) or Bardasis-Schrieffer excitonic modes[37] corresponding to sub-dominant pairing. A theoretical review of THz near-field coupling to these excitations can be found in ref. [38]. Here we briefly comment why these more exotic candidates are unlikely to explain our data. The coupling to Carlson-Goldman modes is expected to be weak because of almost complete charge neutrality of this excitation. Furthermore, disorder is also expected to heavily damp this response[39]. There is no direct optical coupling to pair-breaking excitations or to the Higgs mode of the order parameter[38]. In addition, their energy scale, set by twice the superconducting gap, is larger than our photon energy. The requirements for near-field coupling to excitonic modes also preclude their observation in our experiment. Parity selection rules would forbid an expected *s*- to *d*- transition of the pair wave function. More importantly, Cooper pairs in cuprates are three orders of magnitude smaller than the AFM tip radius used here, so there is a large mismatch in the momentum transfer required to excite these modes. In conclusion, an assessment of near-field coupling mechanisms and energy scales shows that superconductivity-induced collective modes are unlikely to explain our data.

**Supplementary Note 6**
**The spherical dipole model used for AFM-SNOM data analysis**

This section provides details about the model used to calculate the AFM-SNOM signal and the momentum dependence of the probe-sample coupling in Fig. 3 of the main text. We analyzed the tip-sample interaction within the point dipole mode, i.e. by approximating the AFM tip with a sphere whose polarizability is modulated by the presence of the sample[1-4,40], see also Supplementary Figure 5. The effective tip polarizability $\mu$ is given by:

$$\mu(\omega, z_t) = \frac{\alpha}{1 - \alpha \int_0^\infty dq \, r_P(q,\omega) q^2 e^{-2qz_t}} \qquad (13)$$

where $\alpha = a^3 4\pi\varepsilon_0 (\varepsilon_T - 1)/(\varepsilon_T + 2)$ is the bare spherical tip polarizability, $\varepsilon_T$ the dielectric constant of the tip (Tungsten in our case; $\varepsilon_T = -2144 + i \cdot 1015$) and '$a$' its radius. The tip-sample distance $z_t = z_0 + A[1 + \cos(\Omega t)]$ in the tapping mode oscillates with the amplitude '$A$' and has the distance of minimum approach $z_0 \approx a$. The Fresnel reflection coefficient $r_P(q,\omega)$ was discussed in the previous section. Within the dipole model the near-field signal demodulated at the *m*-th harmonic of the tapping frequency is given by:

$$\chi_m(\omega) \cong \frac{1}{T} \int_0^T dt \, \frac{e^{im\Omega t}}{1 - a^3 \int_0^\infty dq \, r_P(q,\omega) q^2 e^{-2q[z_0 + A(1+\cos(\Omega t))]}} \qquad (14)$$

Integration over time of the first term in the Taylor expansion of the denominator in Supplementary Equation (14) for the *m*-th harmonic generates a momentum dependent function $F_m(q, z_0, A)$ given by:

$$F_m(q, z_0, A) = q^2 \cdot e^{-2q(z_0+A)} \cdot I_m(2qA) \approx q^2 \cdot e^{-2qz_0} \qquad (15)$$

The AFM-SNOM momentum form-factor shown with the white dashed line in Fig. 3a of the main text is given by Supplementary Equation (15) for $m = 3$ and for a distance of minimum approach, $z_0$, taken to be equal to the tip radius $a \approx 20$ nm.

The main approximation of the spherical dipole model is the replacement of the actual tip with a spherical particle. The two adjustable parameters of the model are the effective tip polarizability $\propto a^3$, where a is roughly given by the AFM apex radius, and $z_0$, which is the distance of closest approach of the sphere center. The latter parameter should be also of the order of the tip radius. Considering the typical shapes of AFM tips obtained from etched metallic wires, see for example refs. [41,42], as well as the large value of the radiation wavelengths in the THz range, it is clear that the spherical dipole model has its limitations. These shortcomings have quantitative impact on the effective tip polarizability, near-field distribution, field enhancement factors and on effects of retardation. We believe they are at the origin of the quantitative discrepancy between the experimental near-field contrast in Fig. 2 and the model calculation in Fig. 4 of the main text.

These quantitative limitations were also observed in the infrared range. It is expected that the THz regime amplifies them because wavelengths are in the millimeter range. A more realistic tip-shape is going to improve of the agreement with the experiment. Indeed, increasing the effective tip polarizability in the spherical model or the elongation of a prolate spheroid in the quasi-static approximation[43] was found to lead to an increased near-field contrast. These considerations help us understand that the our model can be improved by a more adequate probe geometry. However, because the reasons described above, we believe that a meaningful quantitative agreement can only be achieved by a full electrodynamical treatment taking accurately into account AFM tip shapes and details of the scanning regime.

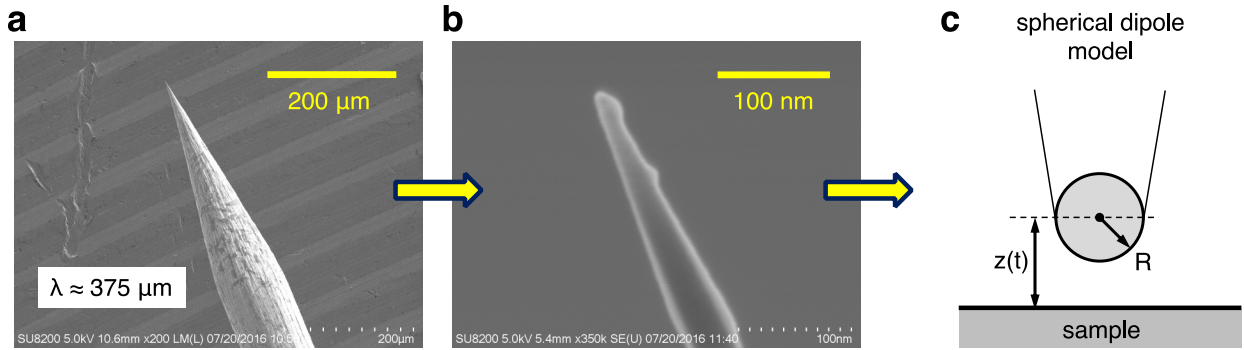

**Supplementary Figure 5. Information about the approximations involved in the spherical dipole model**. (a) Low resolution Scanning Electron Microscope (SEM) image of our typical etched Tungsten AFM tips. The wire diameter is 100 µm. Note the value of the wavelength relative to the tip dimensions. (b) High resolution SEM image showing the tip apex. (c) Parameters of the dipole model.

# Supplementary References